\documentclass[3p,times,twocolumn]{elsarticle}

\usepackage{ecrc}

\volume{00}

\firstpage{1}

\journalname{Nuclear Physics B Proceedings Supplement}

\runauth{CMS Collaboration}

\jid{nuphbp}

\jnltitlelogo{Nuclear Physics B Proceedings Supplement}

\usepackage{amssymb}
\usepackage{amsmath}
\usepackage{xspace}
\usepackage{multirow}

\usepackage[figuresright]{rotating}

\newcommand{\pt}{\ensuremath{p_{\mathrm{T}}}\xspace}
\newcommand{\cm}{\ensuremath{\,\text{cm}}\xspace}
\newcommand{\GeV}{\ensuremath{\,\text{GeV}\xspace}}
\newcommand{\GeVc}{\ensuremath{\,\text{GeV}/c\xspace}}
\newcommand{\GeVcc}{\ensuremath{\mathrm{GeV}/c^2}\xspace}
\newcommand{\TeV}{\ensuremath{\,\text{GeV}\xspace}}
\newcommand{\X}{\ensuremath{\mathrm{X}}\xspace}
\newcommand{\Higgs}{\ensuremath{\mathrm{H}^0\xspace}}
\newcommand{\Higgsdecay}{\ensuremath{\Higgs\rightarrow\X\X}\xspace}

\newcommand{\cPqb}{\ensuremath{b}} 

\providecommand{\PSQ}{\ensuremath{\widetilde{q}\xspace}}
\newcommand{\squark}{\PSQ\xspace}

\newcommand{\sTop}{\ensuremath{\widetilde{t}\xspace}}
\newcommand{\chiz}{\ensuremath{\widetilde{\chi}^{0}}\xspace}
\newcommand{\Pl}{\ensuremath{\ell}\xspace}
\newcommand{\lp}{\ensuremath{\ell^+}\xspace}
\newcommand{\lm}{\ensuremath{\ell^-}\xspace}
\newcommand{\chidecay}{\ensuremath{\chiz\rightarrow\lp\lm\nu}\xspace}
\newcommand{\chidecayH}{\ensuremath{\chiz\rightarrow qq\mu}\xspace}
\newcommand{\sqdecay}{\ensuremath{\squark\rightarrow q\chiz}\xspace}
\newcommand{\BR}{\ensuremath{\mathrm{B}}\xspace}
\newcommand{\PYTHIA} {{\textsc{pythia}}\xspace}
\newcommand{\sigmad}{\ensuremath{\sigma_d}\xspace}
\newcommand{\dxysigma}{\ensuremath{|d_0|/\sigmad}\xspace} 
\newcommand{\dedx}{\ensuremath{\mathrm{d}E/\mathrm{dx}}\xspace}

\begin{document}

\begin{frontmatter}



\dochead{}

\title{Search for long-lived particles at CMS}


\author{Paul Lujan, for the CMS Collaboration}

\address{Princeton University, Department of Physics, Princeton NJ 08544}

\begin{abstract}

The most recent searches for long-lived particles at CMS are presented. Searches for displaced jets,
displaced leptons, displaced stops, and heavy stable charged particles are among those discussed. A variety of
models are constrained by these searches, ranging from hidden valleys to split supersymmetry.

\end{abstract}


\end{frontmatter}


\section{Introduction}
\label{sec:intro}

Many models of new physics predict the existence of new, long-lived particles, which would appear with a very
distinctive experimental signature. Scenarios in which these new particles could arise include supersymmetric
(SUSY) scenarios such as ``split SUSY''~\cite{Hewett:2004nw} or SUSY with very weak 
R-parity violation~\cite{Barbier:2004ez}, ``hidden valley'' models~\cite{Han:2007ae}, and Z' models that
include long-lived neutralinos~\cite{Basso:2008iv}.

The Compact Muon Solenoid (CMS) collaboration~\cite{Chatrchyan:2008aa} has conducted a number of searches for
a variety of different experimental signatures and theoretical models using data collected from proton-proton
collisions at the Large Hadron Collider (LHC). This paper presents an overview of the results of these
searches using 2012 data taken at $\sqrt{s} = 8$~TeV.

There are four analyses discussed here: the first, the ``displaced lepton'' analysis~\cite{pas-exo-12-037},
searches for long-lived neutral particles with a lifetime such that they decay within the CMS detector, but at
a significant displacement from the primary event vertex. The decay products of these particles include a pair
of either electrons or muons. The second, the ``displaced jets'' analysis~\cite{pas-exo-12-038}, considers a
similar model, but for the case where the long-lived particles decay into jets. The third, the ``displaced
SUSY'' analysis~\cite{Khachatryan:2014mea}, considers a model where stops are pair-produced and decay via
$\sTop\sTop \rightarrow be b\mu$. Finally, the ``heavy stable charged particle'' (HSCP)
analysis~\cite{Chatrchyan:2013oca} searches for particles with a large enough lifetime that they do not decay
inside the CMS detector. These particles may have a velocity significantly less than $c$ or a charge
not equal to $\pm e$.

\section{Displaced leptons}
\label{sec:displacedLeptons}

The displaced lepton search is sensitive to a wide class of models that contain long-lived particles decaying
to leptons, but for the context of computing limits, two specific signal models are considered. In the first,
a non-Standard Model (SM) Higgs decays to a long-lived spinless boson X ($H^0 \rightarrow$ XX), which then decays
into a pair of leptons (either X $\rightarrow$ ee or X $\rightarrow \mu\mu$). In the second model, a pair of
squarks is produced in the initial pp collision; each squark then decays into a long-lived neutralino
(\sqdecay). The neutralino then has a R-parity violating decay into a neutrino and two leptons
(\chidecay). Although in these models the long-lived particles are pair-produced and so we would expect to
observe up to two displaced vertices per event, for maximum generality this search only requires one displaced
vertex to be found.

The data is collected using a pair of triggers, one for each channel, requiring either two high-energy
deposits in the electromagnetic calorimeter or two high-momentum tracks in the muon detector. In both cases,
the tracker information is not used in the trigger, as the track reconstruction at the CMS high-level trigger
(HLT) is not necessarily efficient for displaced tracks. In the muon case, the muons are reconstructed without
any primary vertex constraint, and muons from cosmic rays are vetoed by requiring that the three-dimensional
angle between the two muons be less than 2.5 radians.

Simulated signal samples are generated using \PYTHIA to simulate $H^0 \rightarrow$ XX, X $\rightarrow
\ell\ell$ decay for a variety of $H^0$, X masses, and X particle lifetimes corresponding to a transverse decay
length of 2 cm, 20 cm, and 200 cm in the laboratory frame. The \chidecay model is also simulated using \PYTHIA
for a variety of \squark and \chiz masses, with the R-parity violating parameters chosen to give a \chiz
transverse decay length of approximately 20 cm. The signal samples are then reweighted to simulate a variety
of different lifetimes. The background is also simulated using \PYTHIA; the primary background originates from
Drell-Yan production of pairs of leptons, with other contributions from pair production of W or Z bosons,
$t\bar{t}$ production, and QCD multijet events.

\begin{figure}[htp]
  \begin{center}
    \includegraphics[width=0.23\textwidth]{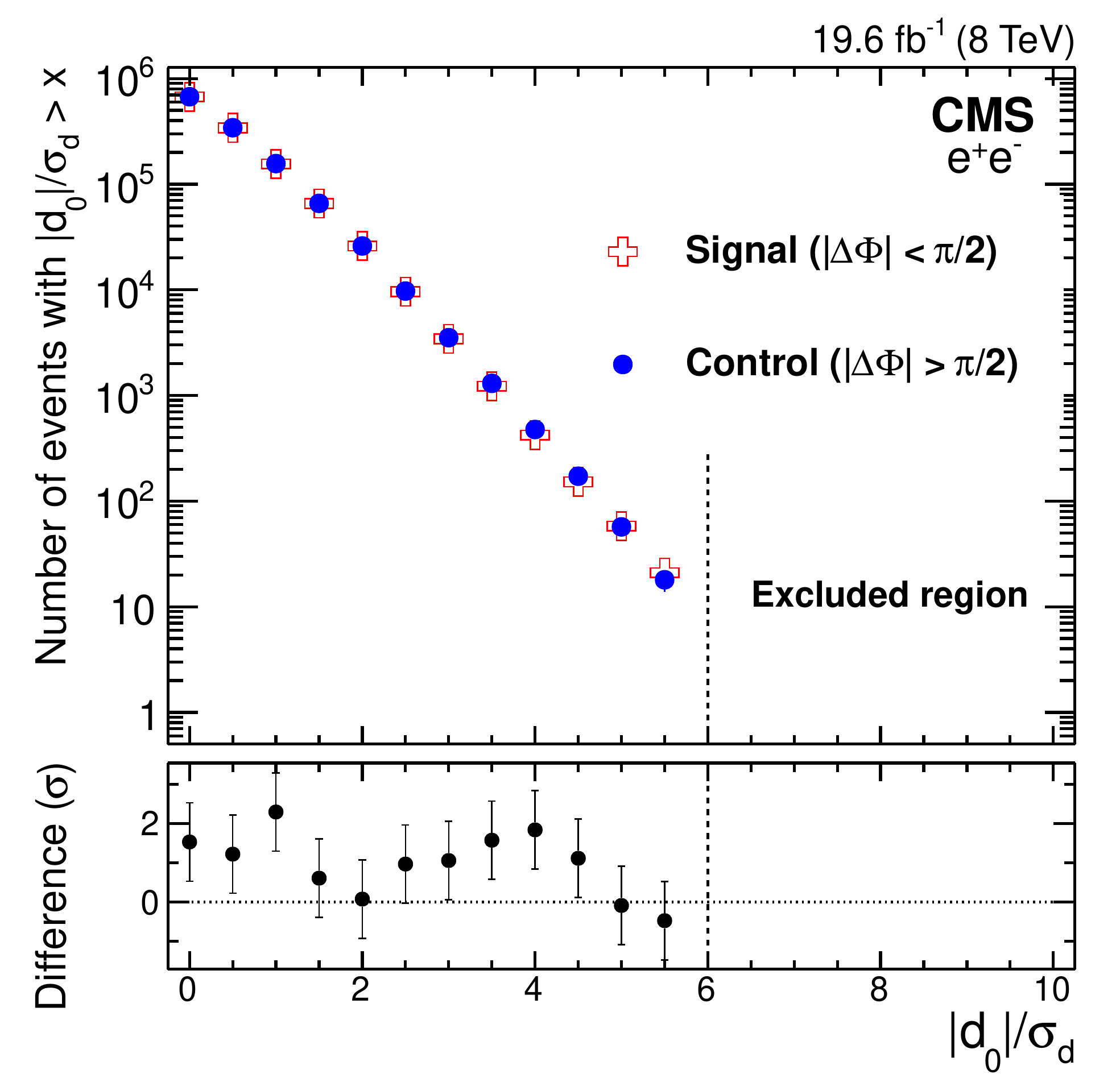}
    \hfill
    \includegraphics[width=0.23\textwidth]{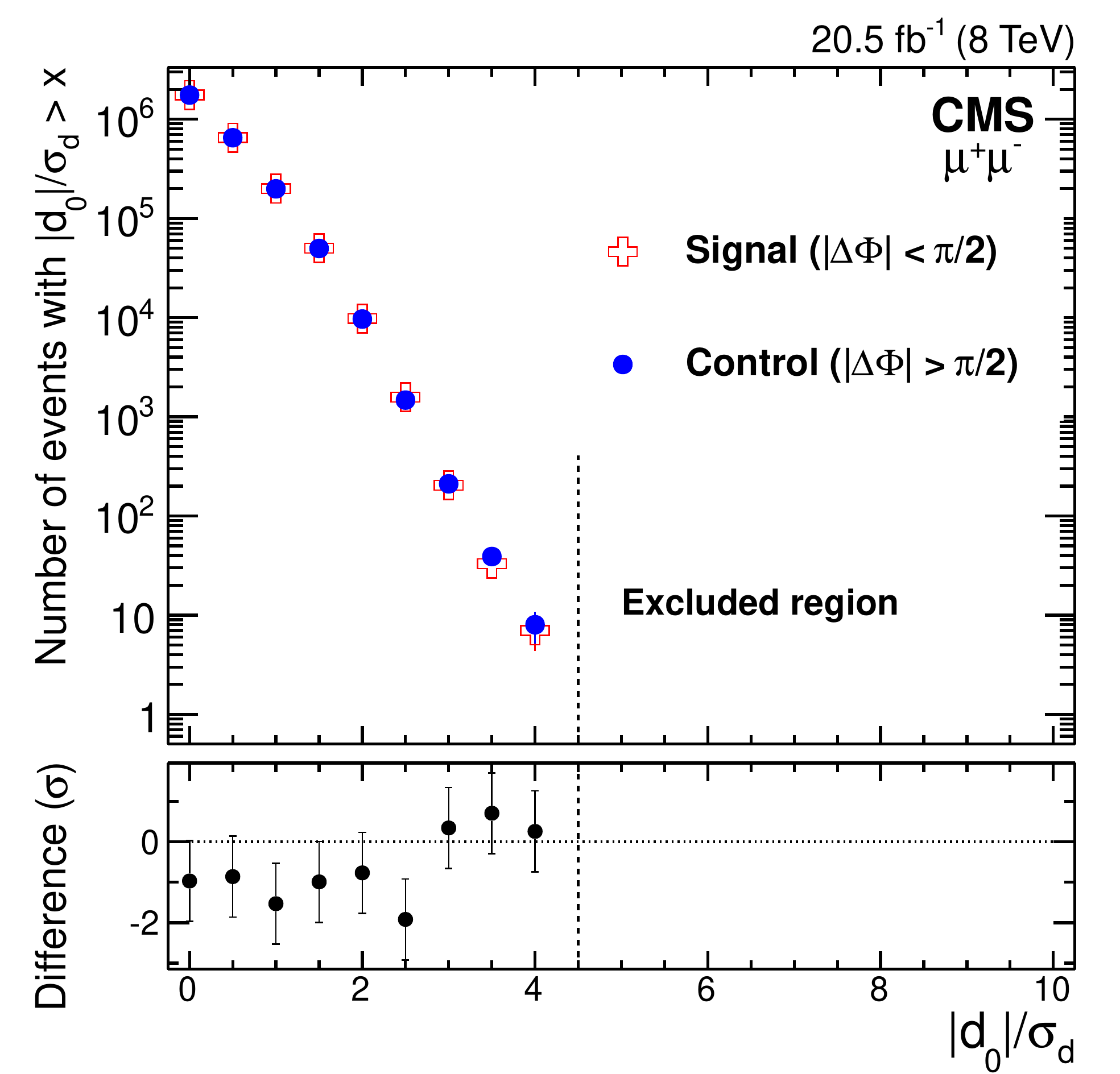}
    \caption{Background estimation for the displaced lepton search. This plot shows the comparison of the
      tail-cumulative distributions (the integral from the given value to infinity) of \dxysigma for data in
      the signal region ($|\Delta\phi| < \pi/2$) and the control region ($|\Delta\phi| > \pi/2$) for the
      dielectron channel (left) and the dimuon channel (right). The high-\dxysigma region is excluded from the
      comparison, as it may potentially contain signal in the signal region. The bottom plots show the
      statistical significance of the difference between the two.}
    \label{fig:DisplacedLeptonsBackground}
  \end{center}
\end{figure}

Long-lived particle candidates are then identified by looking for two isolated, energetic electrons (with $E_T
>$ 40 \GeV) or muons (with $\pt >$ 26 \GeVc) that form a good vertex, and are required to be significantly
displaced from the primary vertex (the significance of the transverse impact parameter $\dxysigma >
12$). The background is estimated using a data-driven method, exploiting the fact that the angle $\Delta\phi$
between the flight direction to the secondary vertex and the dilepton momentum should be close to 0 for a true
long-lived particle event, but randomly distributed for background. Thus, by dividing into a signal region
with $|\Delta\phi| < \pi/2$ and a control region with $|\Delta\phi| > \pi/2$, one can use the number of events
in the latter to estimate the background contribution in the
former. Figure~\ref{fig:DisplacedLeptonsBackground} shows the validity of this method, by comparing the number
of events in the signal and control region as a function of the transverse impact parameter significance
\dxysigma and observing good agreement.

\begin{figure*}[hbtp]
\centering
\includegraphics[width=0.22\textwidth]{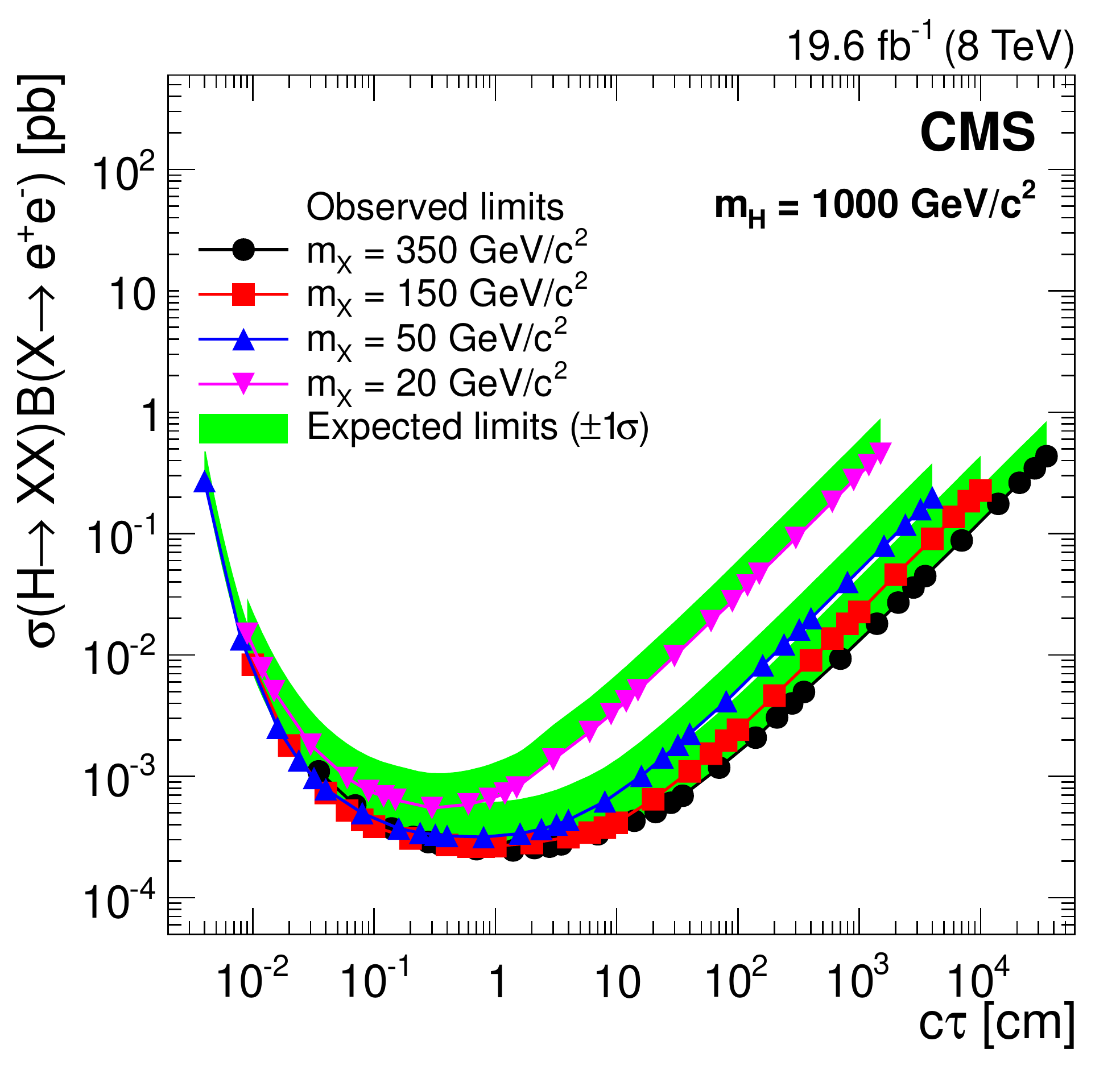}
\includegraphics[width=0.22\textwidth]{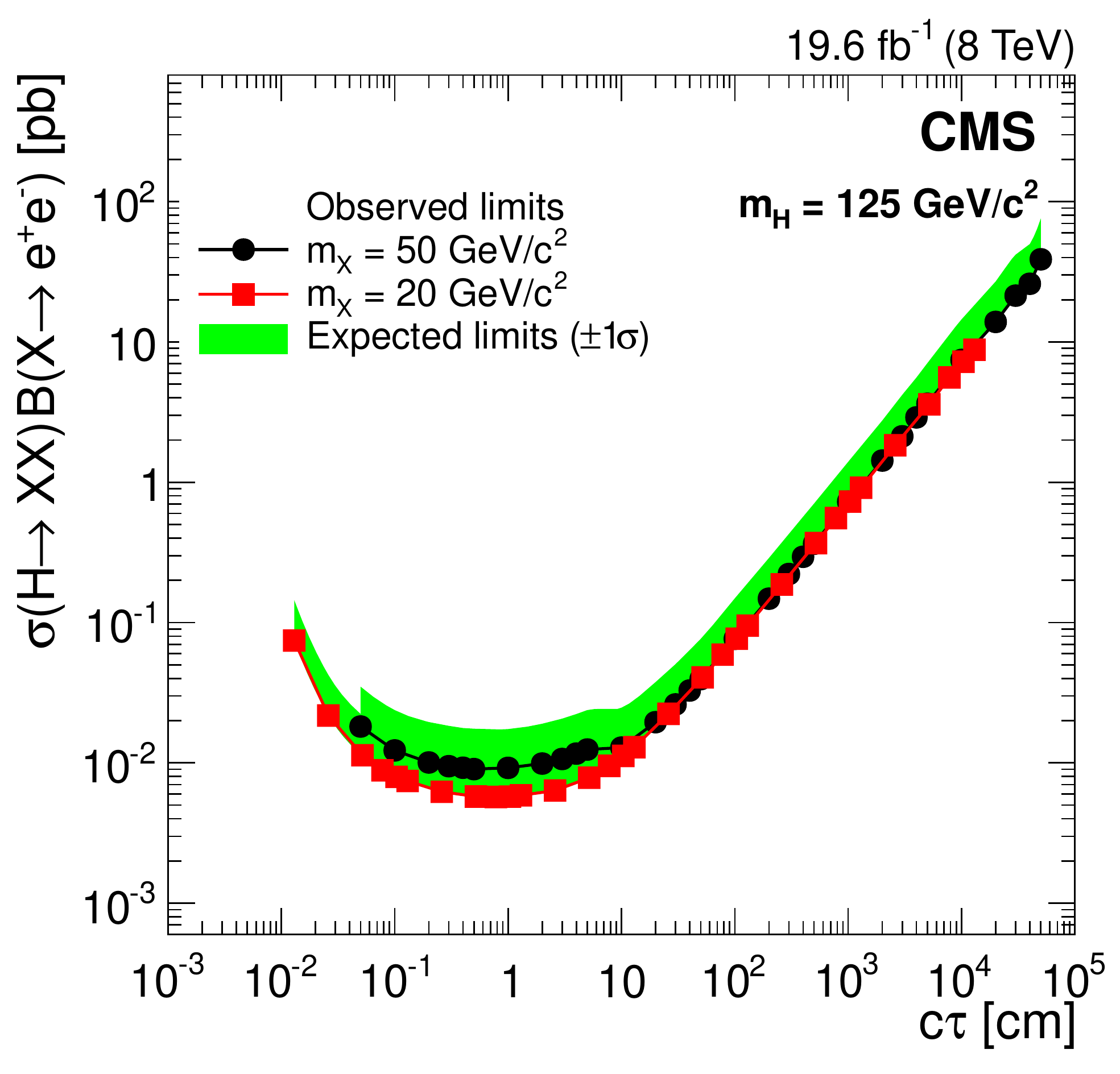}
\includegraphics[width=0.22\textwidth]{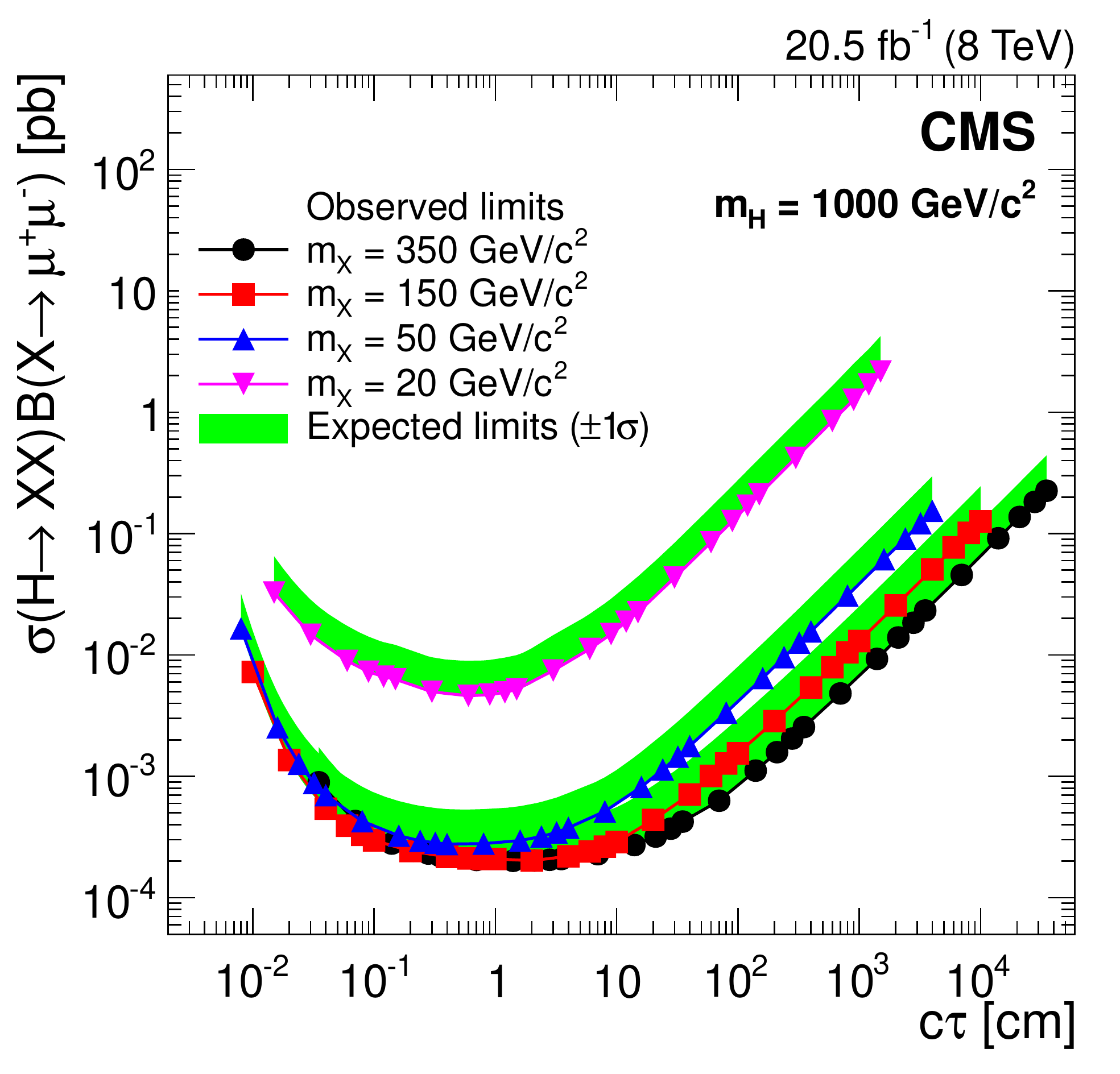}
\includegraphics[width=0.22\textwidth]{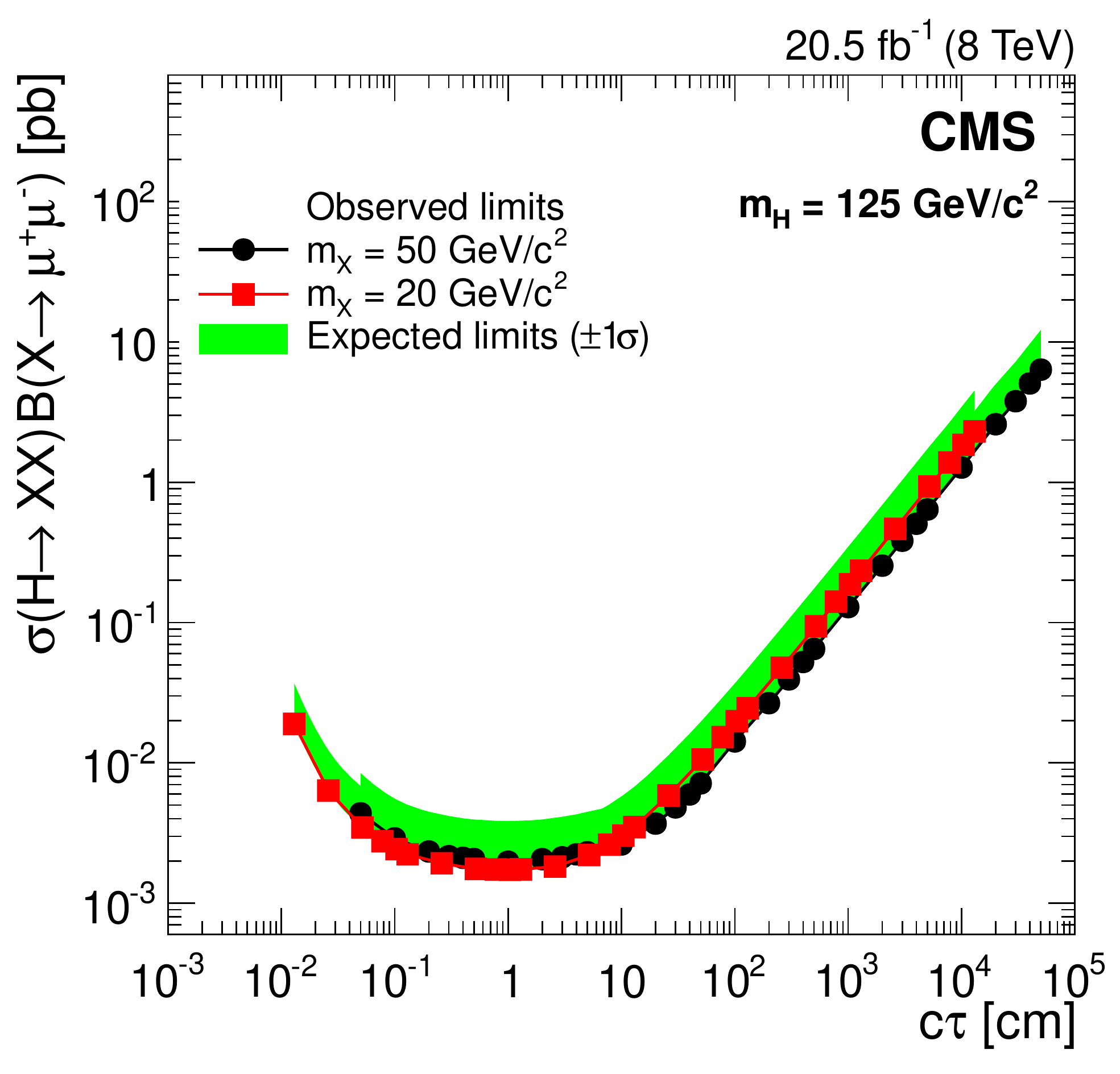}
\caption{The 95\% CL upper limits on $\sigma(\Higgsdecay)\BR(X \rightarrow ee)$ (left two plots) and
  $\sigma(\Higgsdecay)\BR(X \rightarrow \mu\mu)$ (right two plots) for Higgs boson masses of 1000 \GeVcc
  (left) and 125 \GeVcc (right) as a function of the X lifetime. In each plot, results are shown for several
  different \X~boson masses. Shaded bands show the $\pm 1\sigma$ range of variation of the expected 95\% CL
  limits.}
\label{fig:DisplacedLeptonsLimits}
\end{figure*}

The principal systematic uncertainty is from the measurement of the tracking efficiency for displaced
tracks. This is measured by using cosmic ray data to obtain a sample of displaced muons, searching for events
where a muon is reconstructed in the muon chambers in both the upper and lower halves of CMS, and measuring
the resulting efficiency to reconstruct a track in the central tracker. This is repeated with simulated cosmic
ray events and the resulting difference is taken as the systematic. Other notable sources of uncertainty
include the trigger efficiency measurement and the luminosity, with a total uncertainty of approximately
11\%. The estimated background is zero events, and zero events are observed, so limits are
set. Figure~\ref{fig:DisplacedLeptonsLimits} shows some sample limits obtained in the $H^0 \rightarrow$ XX
model for two different $m_{H^0}$ values for a variety of different $m_X$ values and lifetimes.

\section{Displaced jets}
\label{sec:displacedJets}

The displaced jets search looks for a similar class of models as the displaced leptons analysis, but instead
considers the case where the long-lived neutral particle decays hadronically, rather than into
leptons. Specifically, in the $H^0 \rightarrow$ XX model, the X bosons then decay via X $\rightarrow qq$, and
in the long-lived neutralino model, the neutralinos then decay via \chidecayH. As in the displaced lepton
analysis, while these particular models predict two displaced vertices per event, the analysis only requires
one reconstructed displaced vertex, for maximum sensitivity to a variety of models.

The data is collected online using a trigger which requires $H_T > 300$ \GeV, where $H_T$ is the scalar sum of
all jet energies in event, and in addition at least two high-momentum displaced jets ($\pt > 60$ \GeVc). A jet
is identified as displaced in the trigger if no more than two of the tracks associated to the jet have
three-dimensional impact parameters less than 300 $\mu$m, and no more than 15\% of the total jet energy is
carried by tracks with transverse impact parameters less than 500 $\mu$m.

Simulated signal samples are generated using \PYTHIA to simulate $H^0 \rightarrow$ XX, X $\rightarrow qq$
decay for a variety of $H^0$, X masses, and X particle lifetimes corresponding to a transverse decay length of
3 cm, 30 cm, and 300 cm in the laboratory frame. The \chidecayH model is also simulated using \PYTHIA for a
variety of \squark and \chiz masses, with the R-parity violating parameters chosen to give a \chiz transverse
decay length of approximately 20 cm. The signal samples are then reweighted to simulate a variety of different
lifetimes. The only significant background is from QCD multijet events, which are also simulated with \PYTHIA.

\begin{figure}[htp]
  \begin{center}
    \includegraphics[width=0.23\textwidth]{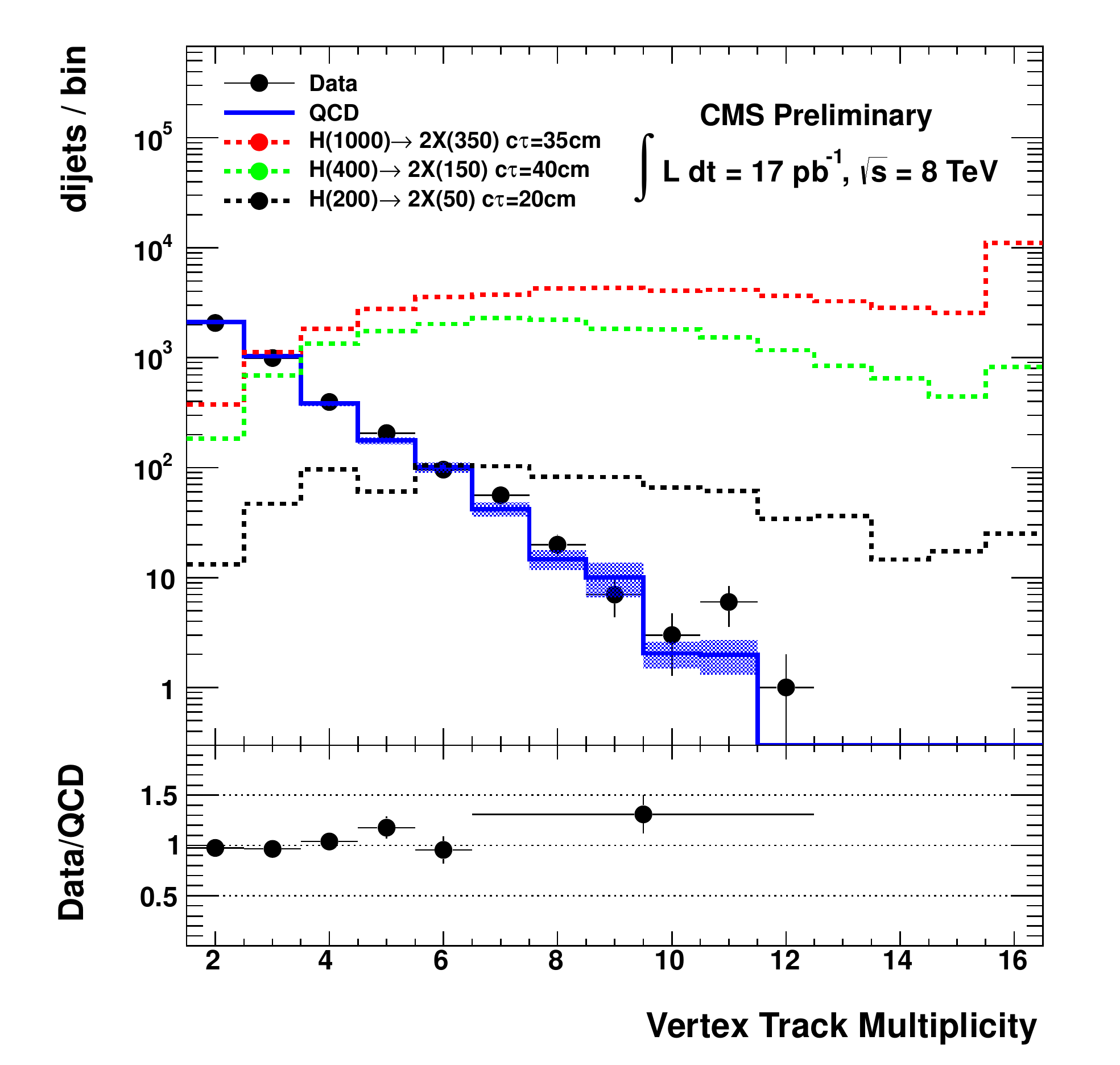}
    \hfill
    \includegraphics[width=0.23\textwidth]{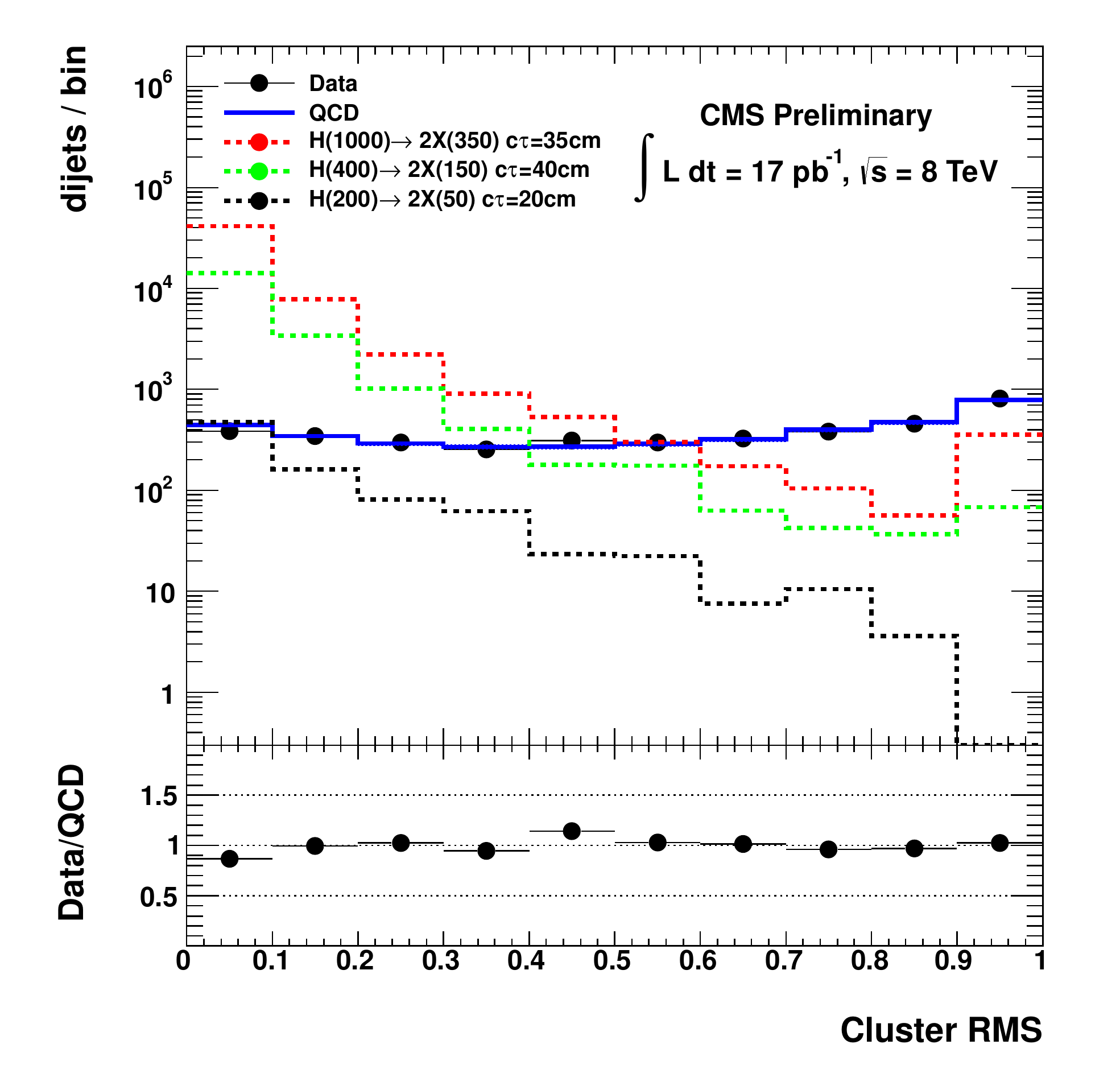}
    \caption{Two of the variables used to build the likelihood discriminant for displaced jets: secondary
      vertex multiplicity (left) and cluster RMS (right). For increased visibility of the signal histograms,
      the signal cross-section is taken to be 10 $\mu$b.}
    \label{fig:DisplacedJetsDiscriminant}
  \end{center}
\end{figure}

Candidate events are then selected by searching for two displaced jets with momentum $> 60$ \GeVc. In
addition, requirements are placed on the number of prompt tracks and fraction of energy carried by prompt
tracks, where a prompt track here is defined as a track with a transverse impact parameter less than 500
$\mu$m. The jets are then required to form a good-quality vertex. Finally, the tracks in the jet are
hierarchically clustered and a likelihood discriminant is built for the jets based on the secondary vertex
track multiplicity, the cluster root-mean-square (RMS), and the fraction of the secondary vertex tracks having
a positive value of the impact parameter. Figure~\ref{fig:DisplacedJetsDiscriminant} shows the distribution of
some of the variables going into the discriminant.

The background estimation is performed using an ABCDEFGH technique, with the eight regions defined by three
selection criteria: the number of prompt tracks and energy fraction of the first jet, the number of prompt
tracks and energy fraction of the second jet, and the likelihood discriminant. Using these, the amount of the
background in the signal region H can be estimated as BCD/A$^2$. The actual values of the selection criteria
are optimized separately for models with lower decay lengths ($L_{xy} <$ 20 cm) and higher ($>$ 20
cm). Table~\ref{tab:DisplacedJetsBackground} shows the final selection criteria and number of expected events.

\begin{table*}[bhtp]
\centering
\begin{tabular}{rcc}
 & $\bf low~\langle L_{xy} \rangle~selection$ & $\bf high~\langle L_{xy} \rangle~selection$ \\
\hline
Number of prompt tracks for each jet & $\leq1$ & $\leq1$ \\
Prompt track energy fraction for each jet & $<0.15$ & $<0.09$ \\
Vertex/Cluster discriminant & $>0.9$ & $>0.8$  \\
\hline
\bf Expected background & $\bf 1.56 \pm 0.25 \pm 0.47$ & $\bf 1.13\pm0.15\pm0.50$ \\
\end{tabular}
\caption{Final selection criteria and the corresponding background expectations for the displaced jet
  analysis. Uncertainties listed are statistical first and systematic second. The low $\ \langle L_{xy}
  \rangle$ selection is optimized for signal models with $\langle L_{xy}\rangle < 20\cm$, while the high
  $\langle L_{xy} \rangle$ selection is optimized for signal models with higher $\langle L_{xy}\rangle$.}
\label{tab:DisplacedJetsBackground}
\end{table*}

The main source of systematic uncertainty on the signal efficiency is from the trigger efficiency, with a
smaller contribution from the track finding efficiency, which is measured using a sample of $K_s^0$
mesons. The total systematic uncertainty is approximately $8-10$\%.

\begin{figure}[hbtp]
\centering
\includegraphics[width=0.23\textwidth]{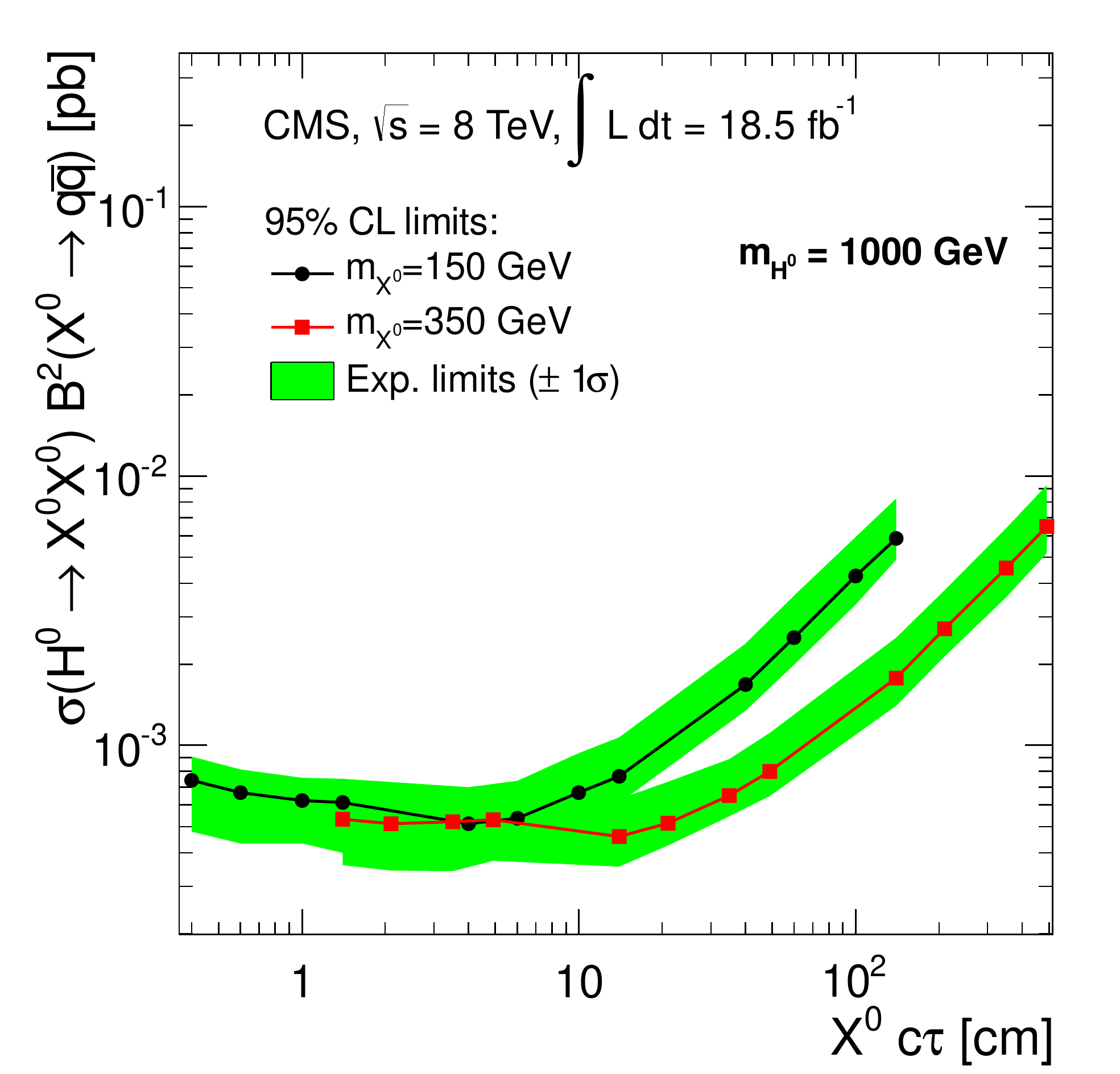}
\includegraphics[width=0.23\textwidth]{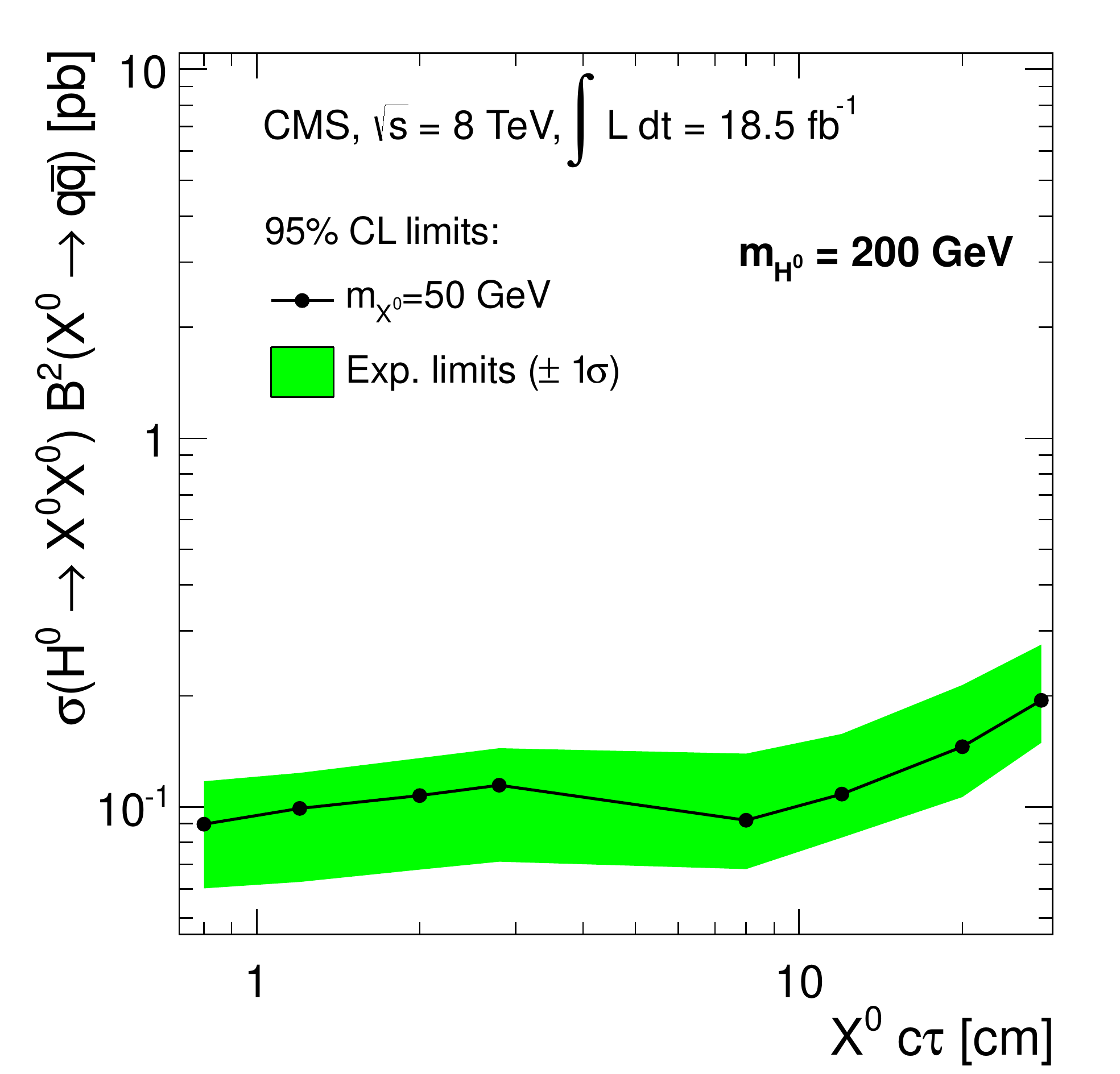}
\caption{The 95\% CL upper limits on $\sigma(\Higgsdecay)\BR^2(X \rightarrow qq)$ for Higgs boson masses of
  1000 \GeVcc (left) and 200 \GeVcc (right) as a function of the X lifetime. In the first plot, results are
  shown for two different \X~boson masses. Shaded bands show the $\pm 1\sigma$ range of variation of the
  expected 95\% CL limits.}
\label{fig:DisplacedJetsLimits}
\end{figure}

In the final results, two events are observed passing the low $L_{xy}$ selection and one passing the high
$L_{xy}$ selection, both consistent with the expected background. Limits are thus set on the signal models;
Figure~\ref{fig:DisplacedJetsLimits} shows a sample of the limits for two different cases.

\section{Displaced supersymmetry}
\label{sec:displacedSUSY}

The displaced SUSY search uses a different search technique to explore a region of parameter space not covered
by previous searches at CMS. Specifically, it considers a SUSY model where long-lived stop quarks are
pair-produced and then undergo an R-violating decay to a $b$ quark and a lepton ($\sTop \rightarrow b
\ell$). The analysis searches for events in the specific channel where one of the leptons is an electron and
the other a muon, thus substantially reducing the background from SM processes. The leptons are required to be
displaced, but they are not required to form a vertex, allowing the search to remain sensitive to a variety of
new physics scenarios.

The data is collected using a trigger requiring a muon with $\pt > 22$ \GeVc and a cluster in the
electromagnetic calorimeter with transverse energy $E_T > 22$ \GeV. \PYTHIA8 is used to generate simulated
signal samples with $pp \rightarrow \sTop \sTop^{*}$. The principal background is from QCD interactions with heavy
flavor (HF), which is modeled using data, with other contributions from $t\bar{t}$ and W/Z production, which are
simulated.

Signal candidate events are identified by looking for an event with a muon and an electron of opposite charge,
both of which are required to be isolated and have $\pt >$ 25 \GeV. The leptons are required to have a
transverse impact parameter $d_0$ less than 2~\cm, so that the tracking efficiency for the leptons remains
high. Three exclusive signal regions are defined: signal region 3 (SR3) requires both leptons to have $d_0 >
0.1$ \cm, signal region 2 (SR2) contains events not in SR3 where both leptons have $d_0 > 0.05$ cm, and signal
region 1 (SR1) contains events not in either of the other two regions where both leptons have $d_0 > 0.02$ cm.

The estimate of background from QCD HF events in the signal region is obtained using an ABCD method with four
regions, containing isolated/non-isolated leptons and same-sign/opposite-sign pairs. The estimated background
from the other sources is estimated from simulation; in order to improve the statistical uncertainty of this
estimation, as the number of simulated events passing all selection criteria may be very small, the
efficiencies for the electron and muon to pass are calculated separately, and then multiplied to find the
final estimate.

The most significant source of systematic uncertainty in the signal efficiency is due to the efficiency for
reconstructing displaced tracks, which is taken as the same as that used in the displaced lepton analysis. The
largest uncertainty in the background estimation is due to the limited sample size for measuring the QCD HF
contribution and amounts to approximately 30\%.

\begin{figure}[hbtp]
\centering
\includegraphics[width=0.48\textwidth]{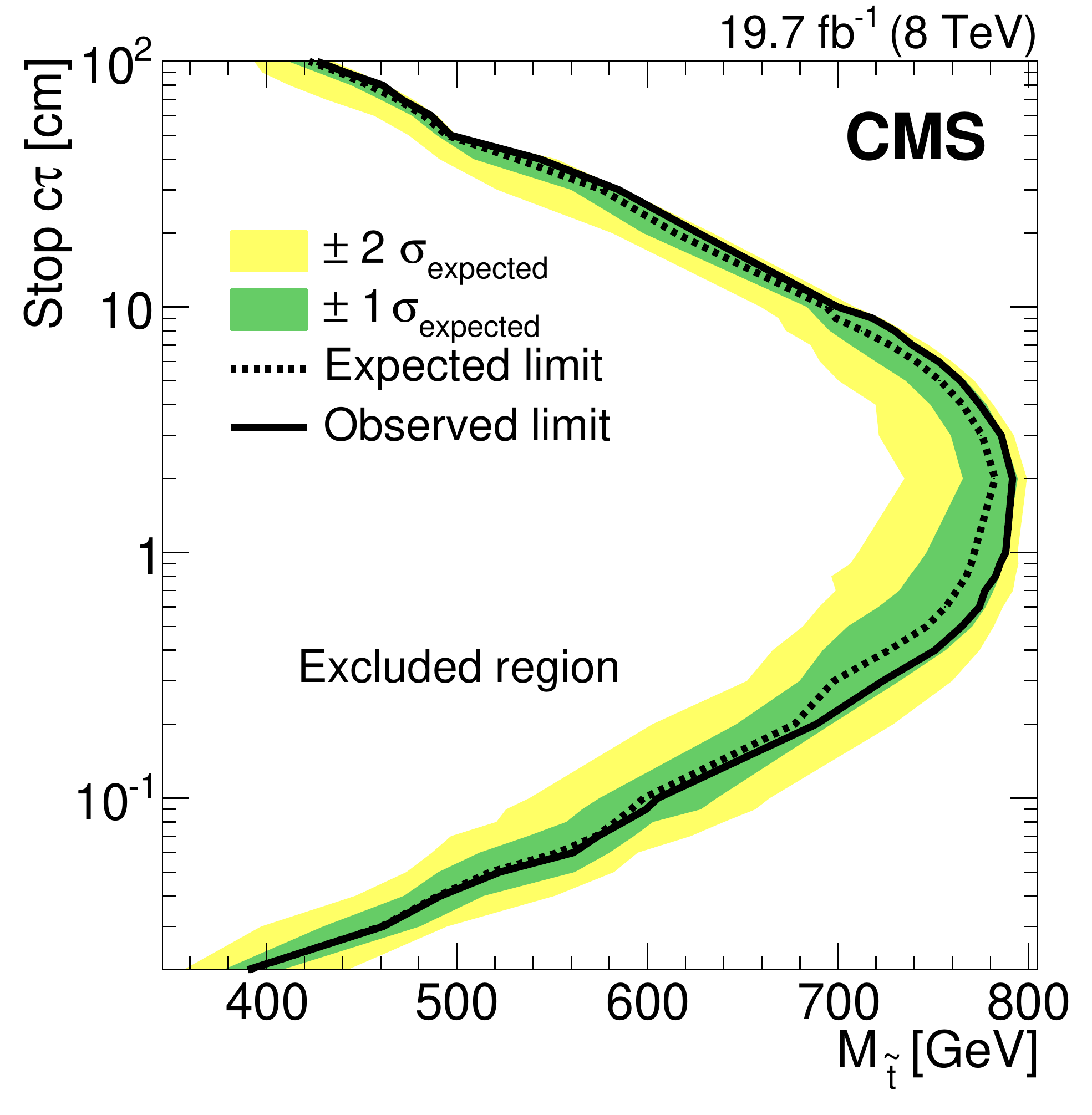}
\caption{Expected and observed $95\%$ CL cross section exclusion contours for top squark pair production 
in the plane of top squark lifetime ($c\tau$) and top squark mass.  
These limits assume a branching fraction of $100\%$ through the RPV vertex 
$\sTop {\rightarrow} \cPqb \Pl$, where the branching fraction to any lepton flavor is equal to $1/3$.
As indicated in the plot, the region to the left of the contours is excluded by this search.}
\label{fig:DisplacedSUSYResults}
\end{figure}

Table~\ref{tab:DisplacedSUSYBackground} shows the expected background and observed number of events in each
search region. In all cases, the observations agree with the expected background, so limits are set on the
stop quark mass as a function of lifetime. Figure~\ref{fig:DisplacedSUSYResults} shows the resulting limits.

\begin{table*}[bhtp]
\centering
\renewcommand{\arraystretch}{1.2}\begin{tabular}{crrr}
\hline
Event source              & \multicolumn{1}{c}{SR1}     & \multicolumn{1}{c}{SR2}                   & \multicolumn{1}{c}{SR3} \\
\hline
Other\ EW                 & $0.65 \pm 0.13 \pm 0.09$    & $(0.89 \pm 0.53 \pm 0.12) \times 10^{-2}$ & ${<}(89 \pm 53 \pm 12) \times 10^{-4}$ \\  
Top quark                 & $0.77 \pm 0.04 \pm 0.08$    & $(1.25 \pm 0.26 \pm 0.12) \times 10^{-2}$ & $(2.4 \pm 1.3 \pm 0.2) \times 10^{-4}$ \\  
Z${\rightarrow}\tau\tau$  & $3.93 \pm 0.42 \pm 0.39$    & $(0.73 \pm 0.73 \pm 0.07) \times 10^{-2}$ & ${<}(73 \pm 73 \pm 7) \times 10^{-4}$ \\  
HF                        & $12.7 \pm 0.2 \pm 3.8$      & $(98 \pm 6 \pm 30) \times 10^{-2}$        & $(340 \pm 110 \pm 100) \times 10^{-4}$ \\  
\hline
Total expected bkgd.      & $18.0 \pm 0.5 \pm 3.8$      & $1.01 \pm 0.06 \pm 0.30$                  & $0.051 \pm 0.015 \pm 0.010$ \\  
\hline
Observed                  & $19$                        & $0$                                       & $0$ \\  
\hline
\end{tabular}
\caption{Numbers of expected and observed events in the three search regions for the displaced SUSY analysis.
  Background and signal expectations are quoted as $N_{\mathrm{exp}} \pm 1\sigma \mathrm{(stat.)} \pm 1\sigma
  \mathrm{(syst.)}$. If the estimated background is zero in a particular search region, the estimate is
  instead taken from the preceding region. Since this should always overestimate the background, it is
  denoted by a preceding ``${<}$''.}
\label{tab:DisplacedSUSYBackground}
\end{table*}

\section{Heavy stable charged particles (HSCP)}
\label{sec:HSCP}

In contrast to the other searches discussed here, the HSCP analysis searches for long-lived particles which
have a sufficiently long lifetime to leave the CMS detector before decaying. These particles can have a speed
$v$ significantly less than $c$, or a charge $Q$ not equal to $\pm e$, or both. The analysis relies on using
the time-of-flight (TOF) of the particles to the muon system and/or the rate of energy loss \dedx in the
tracker for identification of HSCPs.

The search is performed in five separate channels, to account for the wide variety of types of HSCP
possible.
\begin{itemize}
\item The ``tracker+TOF'' analysis searches for singly-charged HSCPs requiring both a track in the inner
tracker and in the muon system.
\item The ``tracker-only'' analysis searches for singly-charged HSCPs requiring only a track in the inner
tracker, to account for the possibility of material interactions causing the HSCPs to become neutral in their
passage through the detector.
\item The ``muon-only'' analysis searches for singly-charged HSCPs requiring only a track in the muon
system.
\item The ``fractionally-charged'' analysis searches for HSCPs with $|Q| < 1e$, requiring a track in the inner
  tracker with a \dedx smaller than that for a SM particle.
\item The ``multiply-charged'' analysis searches for HSCPs with $|Q| > 1e$, requiring tracks in the inner
  tracker and muon system with a much higher \dedx than for the singly-charged analysis.
\end{itemize}

The data is collected using three triggers: a muon trigger requiring a muon with $\pt > 40$ \GeVc, a missing
$E_T$ trigger requiring MET $> 150$ \GeV, and (for the muon analysis) a trigger requiring a muon in the muon
system with $\pt > 70$ \GeVc and MET $> 55$ \GeV. A preselection requiring some general measurements of track
quality is then applied (or quality of the measurement in the muon system in the muon-only case). Signal
samples are simulated using \PYTHIA for models where the HSCP is a long-lived gluino, stop, or stau for a
range of masses. Background is estimated using a purely data-driven approach.

In the tracker+TOF analysis, three variables are used for the principal selection of candidate events: the
track \pt, $1/\beta$, as determined from the time-of-flight to the muon system, and $I_{\textrm{as}}$, a
\dedx-based discriminant for separating SM particle tracks from those with higher \dedx. The other analyses
use a subset of these criteria: the tracker-only analysis uses \pt and $I_{\textrm{as}}$, the muon-only
analysis uses the \pt in the muon system and $1/\beta$, the fractionally-charged analysis uses \pt and
$I_{\textrm{as}}'$, where the prime indicates that the discriminant is now used to identify particles with
lower \dedx, and the multiply-charged analysis uses $1/\beta$ and $I_{\textrm{as}}$ (as the \pt measurement
is not accurate for multiply-charged particles).

These selection criteria are also used to define regions used for an ABCD estimation of the background (or
ABCDEFGH in the case of the tracker+TOF analysis, which has three criteria). The muon-only analysis also
includes an additional estimate of the contribution due to cosmic rays which escape the cosmic ray veto
included in the selection. The principal systematic uncertainties on the signal efficiency measurement are
those on the trigger efficiency, the measurement of \dedx (for all except the muon-only measurement), and the
track momentum reconstruction.

\begin{figure*}[hbtp]
\centering
\includegraphics[width=0.22\textwidth]{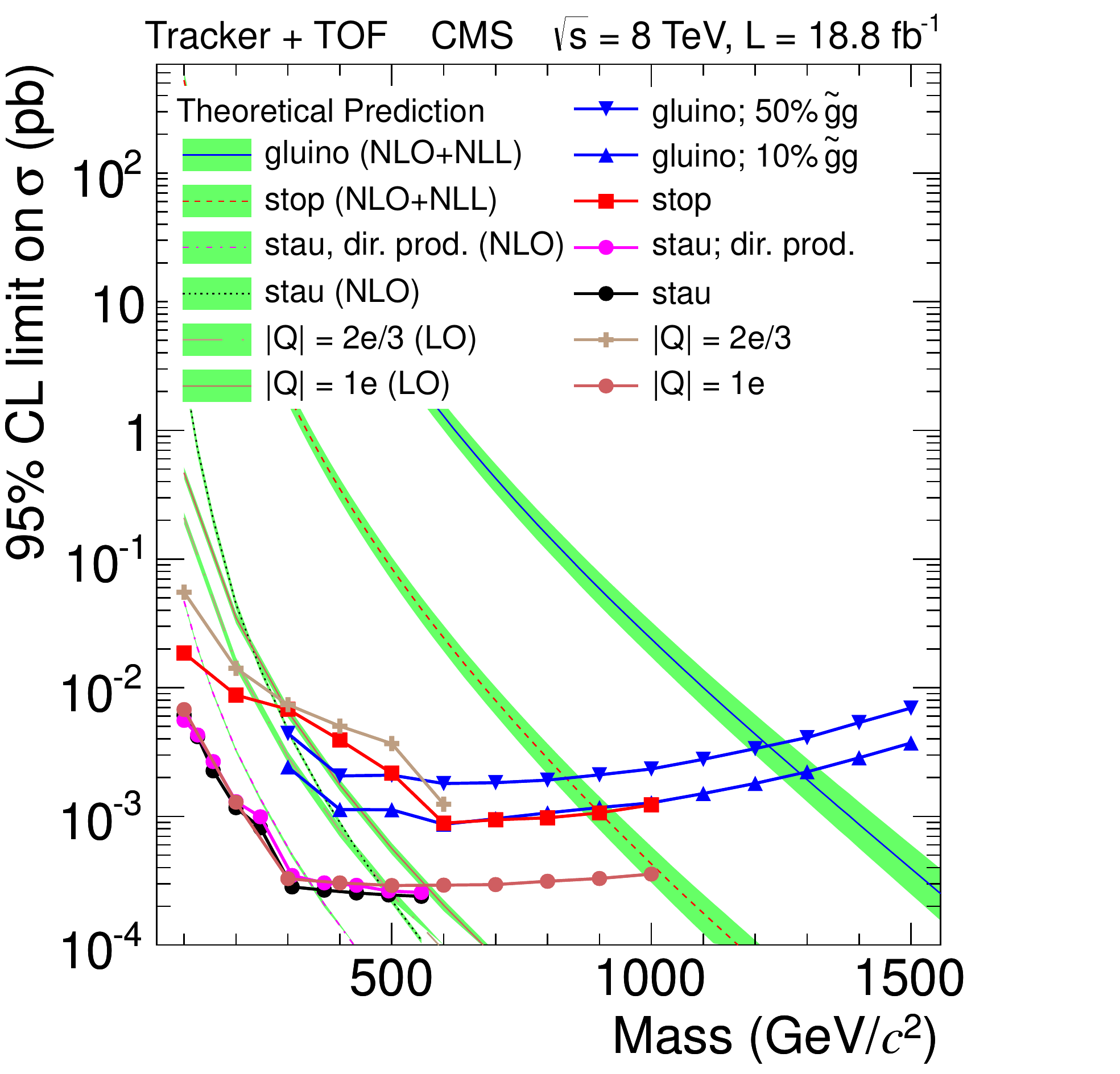}
\includegraphics[width=0.22\textwidth]{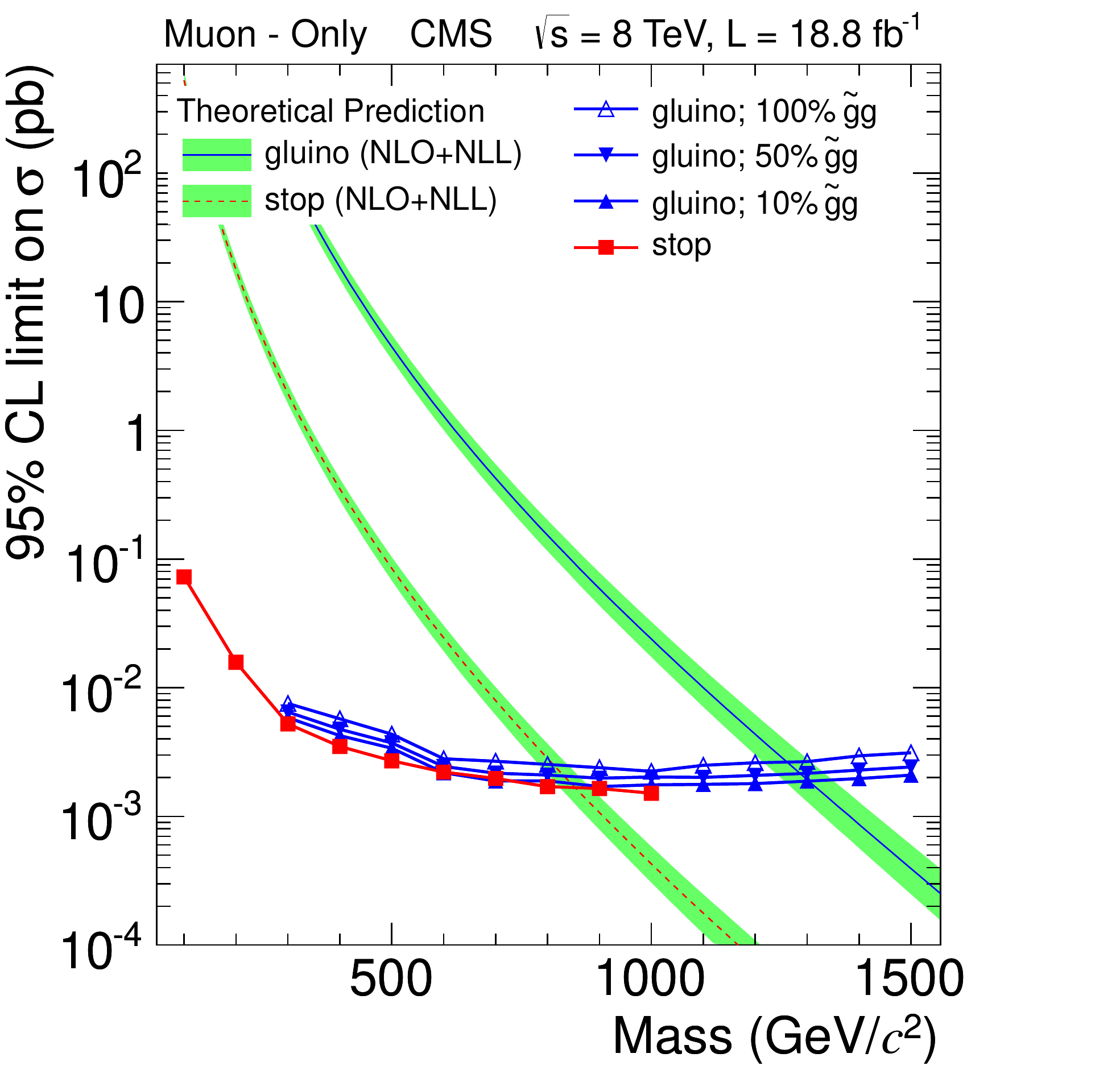}
\includegraphics[width=0.22\textwidth]{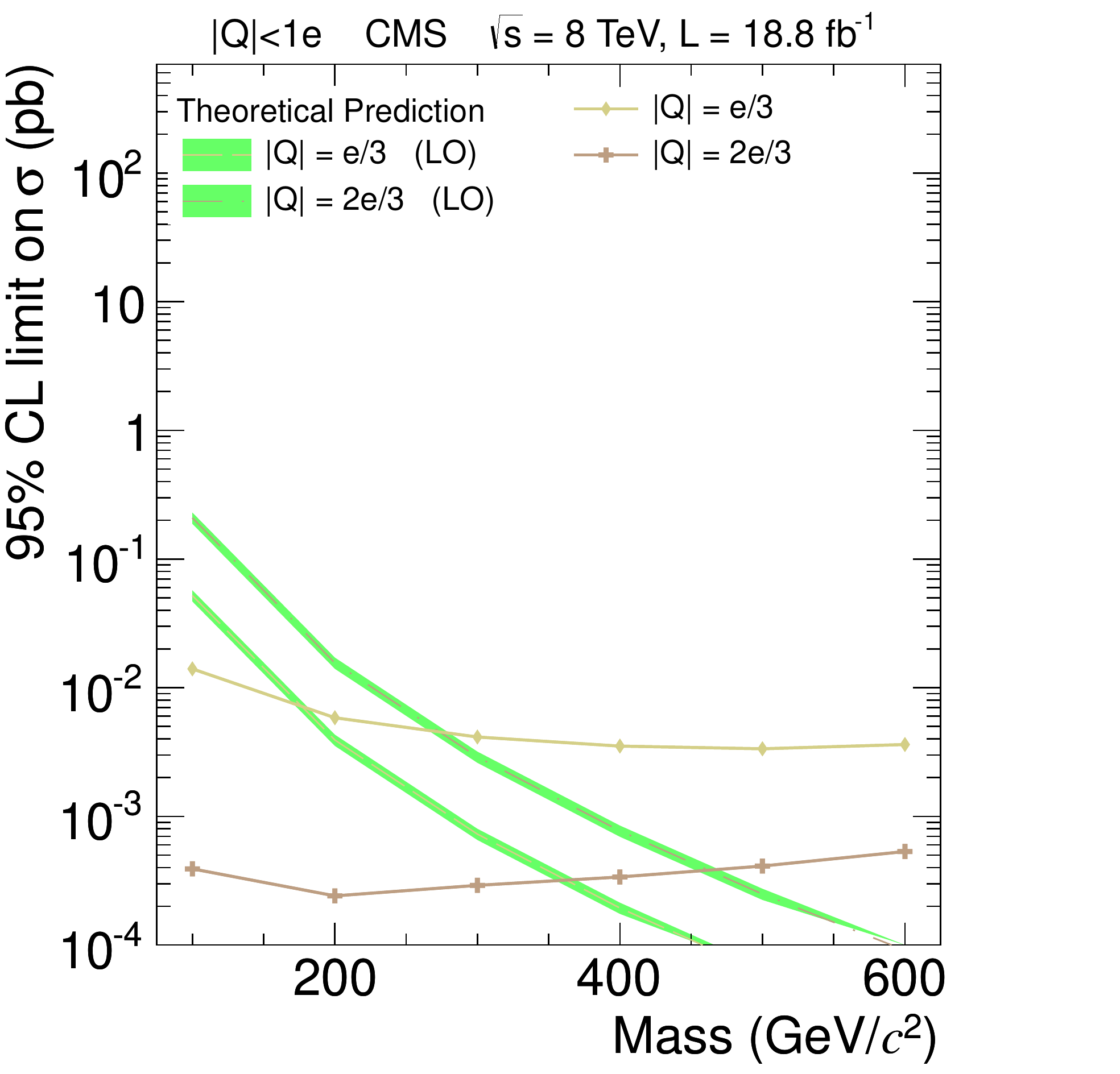}
\includegraphics[width=0.22\textwidth]{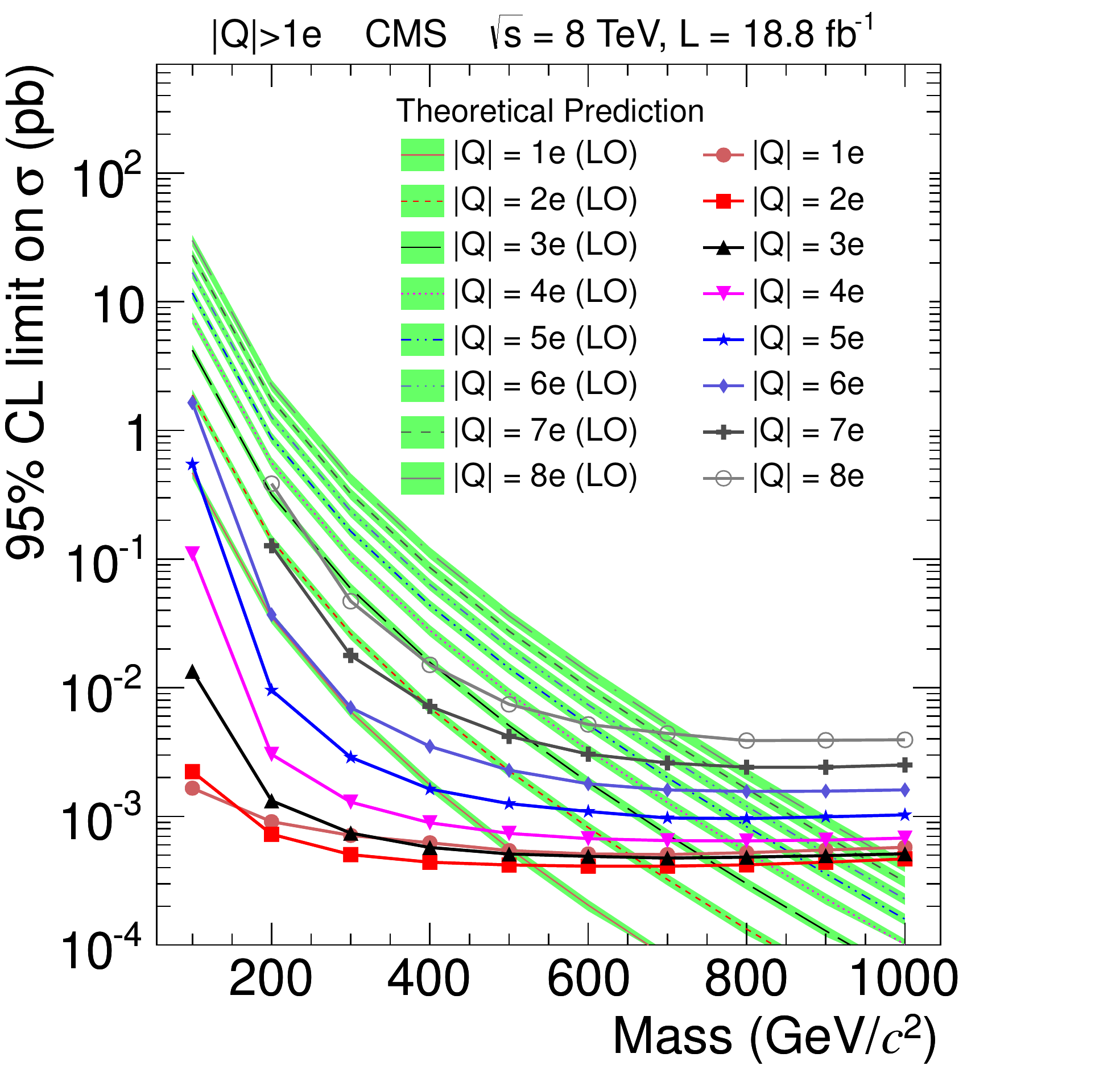}
\caption{Limits for four of the HSCP analysis channels, from left to right: tracker+TOF, muon-only,
  fractionally-charged, and multiply-charged. The different lines correspond to the different signal models
  for which the limits are set.}
\label{fig:HSCPResults}
\end{figure*}

Table~\ref{tab:HSCPBackground} shows the final selection, expected background and observed number of events in
each analysis channel. In all cases, the observations agree with the expected background, so limits are set
for the three signal models considered. Figure~\ref{fig:HSCPResults} shows some example results.

\begin{table*}
 \begin{center}
  \small
 \begin{tabular}{|p{1.8cm}|c|c|c|c|cc|} \hline
                             & \multicolumn{4}{c|}{Selection criteria}              & \multicolumn{2}{c|}{Number of events, $\sqrt{s}=8$\TeV} \\ \hline
                             & \pt                   & \multirow{2}{*}{$I_{as}^{(\prime)}$} & \multirow{2}{*}{$1/\beta$} & Mass    & \multirow{2}{*}{Pred.} & \multirow{2}{*}{Obs.}             \\
                             & (\GeVc)                 &                            &                            & (\GeVcc)&                & \\ \hline
   \multirow{4}{*}{\hspace{-0.1cm}Tracker-only  }& \multirow{4}{*}{${>}70$} & \multirow{4}{*}{${>}0.4$}   & \multirow{4}{*}{$-$}       & ${>}0$ & $33\pm7$   & $41$ \\
                             &                         &                            &                            & ${>}100$ & $26\pm5$   & $29$ \\
                             &                         &                            &                            & ${>}200$ & $3.1\pm0.6$    & $3$  \\
                             &                         &                            &                            & ${>}300$ & $0.55\pm0.11$  & $1$  \\
                             &                         &                            &                            & ${>}400$ & $0.15\pm0.03$  & $0$ \\ \hline
  \multirow{4}{*}{\hspace{-0.1cm}Tracker+TOF\ }& \multirow{4}{*}{${>}70$} & \multirow{4}{*}{${>}0.125$} & \multirow{4}{*}{${>}1.225$} & ${>}0$ & $44\pm9$   & $42$ \\
                             &                         &                            &                            & ${>}100$ & $5.6\pm1.1$    & $7$  \\
                             &                         &                            &                            & ${>}200$ & $0.56\pm0.11$  & $0$  \\
                             &                         &                            &                            & ${>}300$ & $0.090\pm0.02$ & $0$ \\ \hline
   \hspace{-0.1cm}Muon-only  & ${>}230$                 &  $-$                       & ${>}1.40$                   &    $-$  & $6\pm3$    & $3$ \\ \hline
   \hspace{-0.1cm}$|Q|>1e$   &           $-$           &       ${>}0.500$            &        ${>}1.200 $          &    $-$  & $0.52\pm 0.11$ & $1$ \\ \hline
   \hspace{-0.1cm}$|Q|<1e$   & ${>}125$                 & ${>}0.275$                   &            $-$             &    $-$  & $1.0\pm0.2$  & $0$ \\ \hline
 \end{tabular}
 \normalsize
 \end{center}
 \caption{The final selections used for the different channels of the HSCP analysis, the predicted
   background in each case, and the observed number of events. The uncertainties include both statistical and
   systematic contributions.
   \label{tab:HSCPBackground}}
\end{table*}

\begin{figure}[hbtp]
\centering
\includegraphics[width=0.33\textwidth]{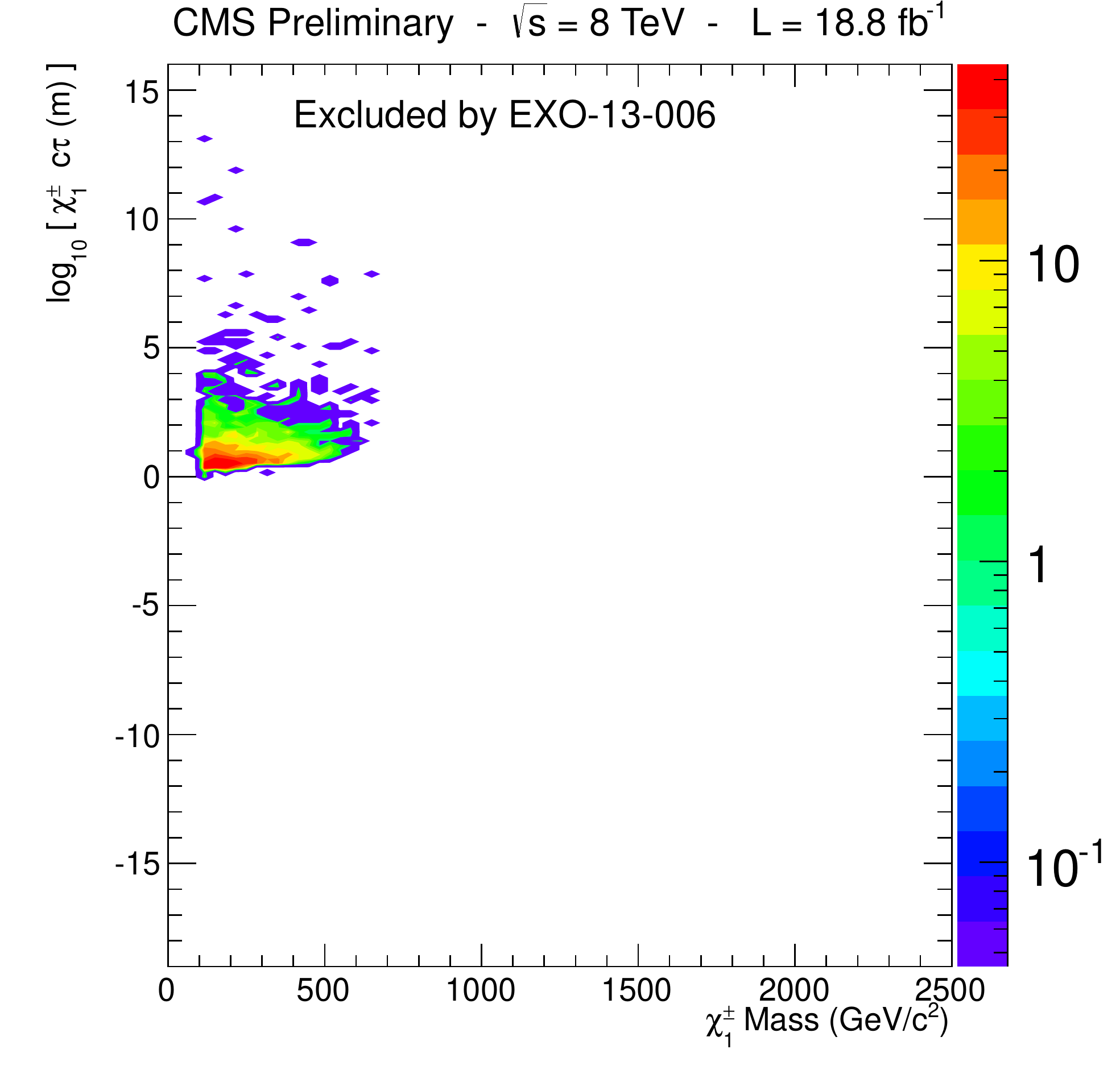}
\caption{Parameter points in the pMSSM that can be excluded by reinterpretation of the HSCP search as
  described in the text.}
\label{fig:HSCPReinterpretation}
\end{figure}

These results can also be reinterpreted in the context of other signal models.~\cite{pas-exo-13-006} This is
performed by constructing efficiency maps for reconstructing these particles as a function of $\beta$ and
$\eta$ in bins of $\pt$. Then, one can calculate the efficiency for any given signal model simply by
calculating the predicted kinematics for that model and applying the efficiency
maps. Figure~\ref{fig:HSCPReinterpretation} shows this technique used to show points in the phase space of the
phenomenological minimal supersymmetric model (pMSSM) that can be excluded by this analysis.

\nocite{*}
\bibliographystyle{elsarticle-num}
\bibliography{LongLivedParticlesCMS}







\end{document}